\documentclass[12pt]{article}
\usepackage{times}
\usepackage{geometry}
\usepackage{amssymb}     
\usepackage{ragged2e}
\usepackage[T1]{fontenc}
\usepackage[figure,figure*]{}
\usepackage{graphicx}
\usepackage{ulem} 
\usepackage[T1]{fontenc}
\usepackage[figure,figure*]{}
\usepackage{graphicx}
\usepackage{ulem} 
\usepackage{caption}
\usepackage{subcaption}

\usepackage{color}
\usepackage{lineno}
\usepackage{wrapfig}
\usepackage{subcaption}
\usepackage{float}

\usepackage[utf8]{inputenc}
\usepackage{enumitem,amssymb}
\usepackage{ragged2e}

\newlist{thematic}{itemize}{8}
\setlist[thematic]{label=$\square$}
\usepackage{pifont}

\newcommand{\otwo}{O$_2$}
\newcommand{\ozone}{O$_3$}
\newcommand{\methane}{CH$_4$}
\newcommand{\water}{H$_2$O}
\newcommand{\cotwo}{CO$_2$}

\begin{document}
\raggedright
\huge
Astro2020 Science White Paper \linebreak

Detecting Earth-like Biosignatures on Rocky Exoplanets around Nearby Stars with Ground-based Extremely Large Telescopes
\linebreak
\normalsize

\noindent \textbf{Thematic Areas:} \hspace*{55pt} $\boxtimes$ Planetary Systems \hspace*{10pt} $\square$ Star and Planet Formation \hspace*{20pt}\linebreak
$\square$ Formation and Evolution of Compact Objects \hspace*{31pt} $\square$ Cosmology and Fundamental Physics \linebreak
  $\square$  Stars and Stellar Evolution \hspace*{1pt} $\square$ Resolved Stellar Populations and their Environments \hspace*{40pt} \linebreak
  $\square$    Galaxy Evolution   \hspace*{45pt} $\square$             Multi-Messenger Astronomy and Astrophysics \hspace*{65pt} \linebreak
  
\textbf{Principal Author:}

Name: Mercedes L\'opez-Morales	
 \linebreak						
Institution: Center for Astrophysics | Harvard \& Smithsonian 
 \linebreak
Email: mlopez-morales@cfa.harvard.edu
 \linebreak
Phone: +1.617.496.7818
 \linebreak
 
\textbf{Co-authors:} Thayne Currie (NASA-Ames/Subaru Telescope),  Johanna Teske (Carnegie Observatories), Eric Gaidos (U. of Hawaii), Eliza Kempton (U. of Maryland), Jared Males (U. of Arizona), Nikole Lewis (Cornell University), Benjamin V. Rackham (U. of Arizona), Sagi Ben-Ami (CfA), Jayne Birkby (U. of Amsterdam, Netherlands), David Charbonneau (CfA), Laird Close (U. of Arizona), Jeff Crane (Carnegie Observatories), Courtney Dressing (U. of California Berkeley), Cynthia Froning (U. of Texas at Austin),Yasuhiro Hasegawa (NASA-JPL/Caltech), Quinn Konopacky (U. of California San Diego), Ravi K. Kopparapu (NASA-GSFC), Dimitri Mawet (California Institute of Technology), Bertrand Mennesson (Caltech/NASA-JPL), Ramses Ramirez (ELSI/Tokyo Institute of Technology), Deno Stelter (U. of California Santa Cruz), Andrew Szentgyorgyi (CfA), Ji Wang (Ohio State University)
  \linebreak
  
\textbf{Co-signers:} Munazza Alam (CfA), Karen Collins (CfA), Andrea Dupree (CfA), Julien Girard (STScI), Rapha\"elle Haywood (CfA), 
Margarita Karovska (CfA), James Kirk (CfA), Amit Levi (CfA), Christian Marois (NRC-Herzberg), Chima McGruder (CfA), Chris Packman (U. of Texas San Antonio), Paul Rimmer (U. of Cambridge), Sarah Rugheimer (U. of Oxford, UK), Surangkhana Rukdee (CfA)
  \linebreak

\justify
\textbf{Abstract}
As we begin to discover rocky planets in the habitable zone of nearby stars with missions like TESS and CHEOPS, we will need quick advancements on instrumentation and observational techniques that will enable us to answer key science questions, such as {\it What are the atmospheric characteristics of habitable zone rocky planets?} {\it How common are Earth-like biosignatures in rocky planets?} {\it How similar or dissimilar
are those planets to Earth?} 
Over the next decade we expect to have discovered several Earth-analog candidates, but we will not have the tools to study the atmospheres of all of them in detail.
Ground-based ELTs can identify biosignatures in the spectra of these candidate exo-Earths and understand how the planets' atmospheres compare to the Earth at different epochs.   Transit spectroscopy, high-resolution spectroscopy, and reflected-light direct imaging on ELTs can identify multiple biosignatures for habitable zone, rocky planets around M stars at optical to near-infrared wavelengths.   Thermal infrared direct imaging can detect habitable zone, rocky planets around AFGK stars, identifying ozone and motivating reflected-light follow-up observations with NASA missions like HabEx/LUVOIR.
Therefore, we recommend that the Astro2020 Decadal Survey Committee support: (1) the search for Earth-like biosignatures on rocky planets around nearby stars as a key science case; (2) the construction over the next decade of ground-based Extremely Large Telecopes (ELTs), which will provide the large aperture and spatial resolution necessary to start revealing the atmospheres of Earth-analogues around nearby stars; (3) the development of instrumentation that optimizes the detection of biosignatures; and (4) the generation of accurate line lists for potential biosignature gases, which are needed as model templates to detect those molecules.

\pagebreak
\vspace{3mm}
\noindent {\bf 1.~The case for Earth-like Biosignatures}
\vspace{1mm}

{\it Are we alone in the Universe?} {\it Are there other planets like Earth?} These are centuries-old questions, which we may soon be poised to answer with the advent of technology to detect rocky planets like Earth and to search their atmospheres for biosignatures, atmospheric gases indicative of biological activity. Though many molecules can be a potential biosignature (see e.g. Seager et al. 2016), given our current one-planet-with-life sample, we can start by looking for a combination of molecular oxygen (\otwo), water (\water), ozone (\ozone), methane (\methane), and carbon dioxide (\cotwo) as habitability indicators. \otwo{} is considered a key biosignature gas, particularly when found in the disequilibrium combination of \otwo\ and \methane\ (Lovelock 1965, Sagan et al. 1993).  \ozone\ is a photochemical derivative of O$_2$, and \water\ and \cotwo\ are important greenhouse gases that maintain habitable conditions (e.g.~Kaltenegger et al. 2007). Combined in the right proportions in the atmosphere of an exoplanet, these molecules might indicate the presence of life.   There are also suggested abiotic pathways for \otwo, \ozone,  and \methane\  on Earth-like planets (e.g.,~Domagal-Goldman et al. 2014, Wordsworth \& Pierrehumbert 2014, Luger \& Barnes 2015, Kietavainen $\&$ Purkamo 2015), but the appearance of these gases on Earth’s atmosphere is strongly linked to the emergence of oxygenic photosynthesis by cyanobacteria-like organisms (see, e.g.,~Schwieterman et al. 2018, and references therein), and methanogenic organisms (e.g.~Lessner 2009). This could also be the case on other planets.  

\begin{figure*}[h!]
\centering
\includegraphics[angle=0, scale=0.485, trim=0cm 0cm 0cm 0cm,clip=true]{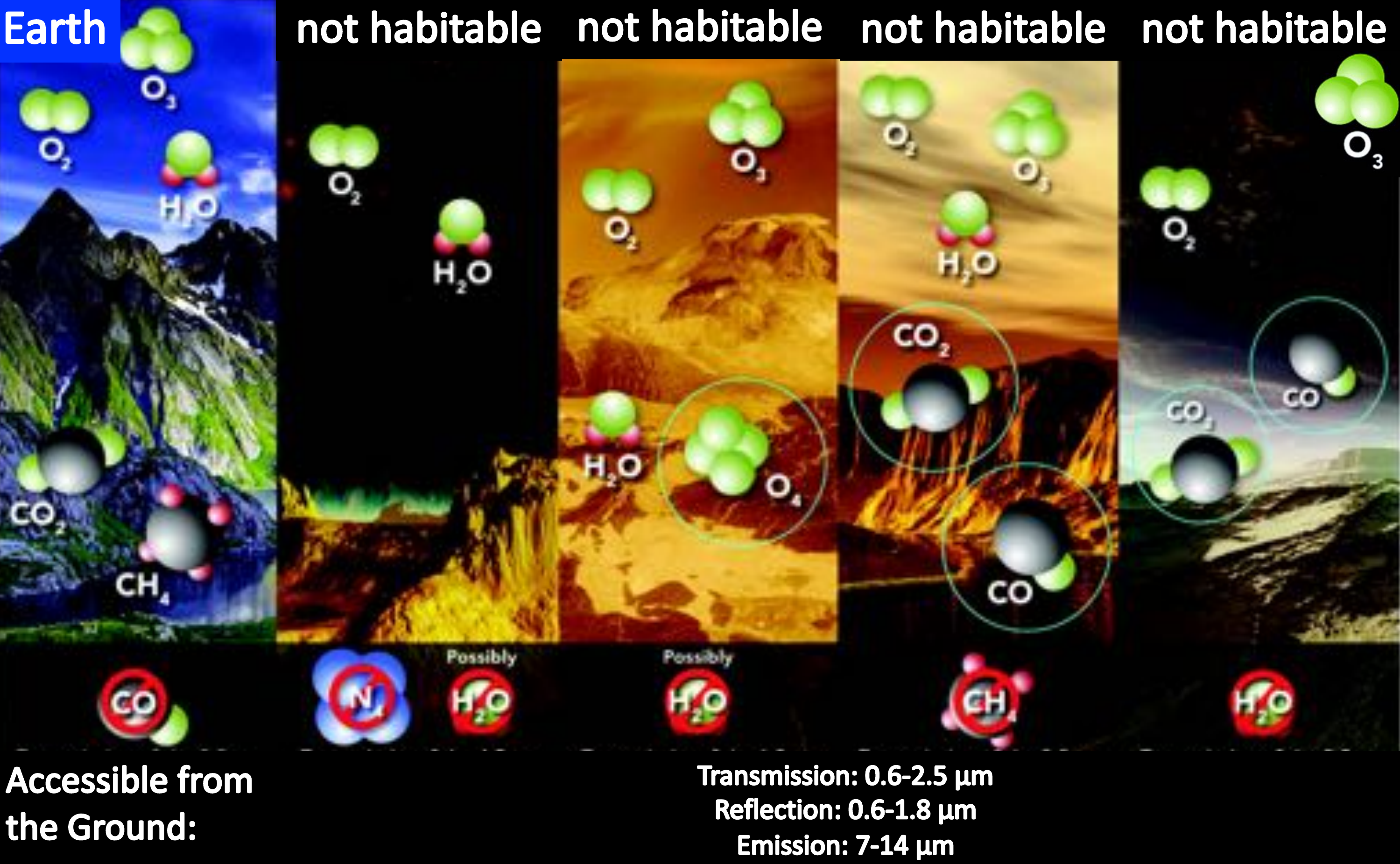}
\vspace{-0.1in}
\caption{Predicted atmospheric composition scenarios of habitable and not habitable Earth-analogues. Molecules circled in cyan, if detected, would help reveal mechanisms that produce ${\rm O_2}$ abiotically. Molecules highlighted at the bottom with the forbidden sign would not be in the atmosphere of the planet in these scenarios. Figure adapted from Fig.~2 in Meadows et al. 2018. See their paper for more details.\label{fig:atmtypes}}
\vspace{-0.2in}
\end{figure*}

\begin{figure}[h]
 \begin{subfigure}[h]{0.5\textwidth}
\includegraphics[angle=0, width=0.99\linewidth, trim=0cm 0cm 0cm 0.5cm,clip=true]{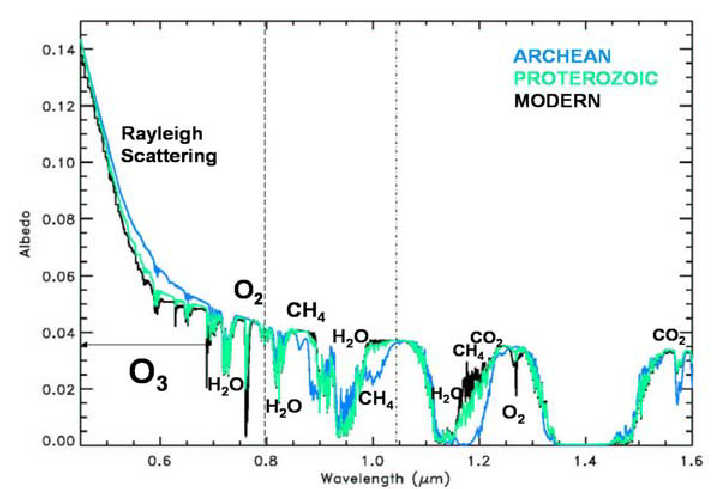} 
\end{subfigure}
\begin{subfigure}[h]{0.5\textwidth}
\includegraphics[angle=0, width=0.97\linewidth,  trim=0cm 0cm 0cm 0cm,clip=true]{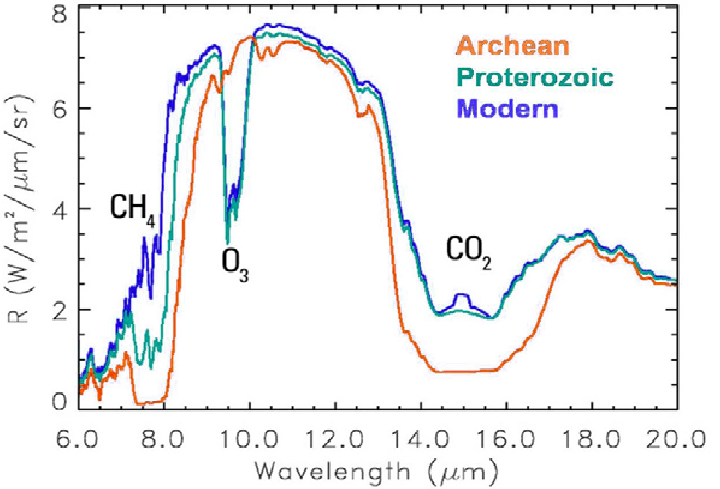}
\end{subfigure}
\vspace{-0.175in}
 \caption{${\it Left}$ -- 0.5 to 1.6 $\mu$m  model reflectance spectra of a cloudless Earth during the late Archean (3.2--2.3\,Gya), Proterozoic (2.3--0.8\,Gya) and Modern (present day). ${\it Right }$ -- 6 to 20 $\mu$m model emission spectra of cloudless Earth during the same eons. ${\rm O_2}$ and ${\rm O_3}$ absorption bands become prominent after the Archean, when fossil records reveal the onset of widespread photosynthetic oxygenation on Earth's surface. ${\rm CH_4}$ and ${\rm CO_2}$ features weaken around that same time. Figures from Meadows (2006). \label{fig:archean}}
 \vspace{-0.2in}
\end{figure}

Observations of \water, \cotwo, \otwo, \ozone, and \methane\ can reveal the atmospheric diversity of rocky planets. The absence of any of these gases may imply departures from Earth-like and habitable conditions (Figure~\ref{fig:atmtypes}). Such observations can also be compared to atmospheric models of Earth at different geologic eons (Figure~\ref{fig:archean}; Meadows 2006), constraining how the planets are similar to an earlier version of Earth.

\vspace{3mm}
\noindent {\bf 2.~The Sample of Targets and their Observations with ELTs}
\vspace{1mm}

Statistics from the Kepler mission reveal that about 50\% of M dwarf stars have at least one small, habitable zone (HZ) planet (Dressing \& Charbonneau 2015). Results for more massive stars are currently unknown, but establishing that occurrence rate---especially for Sun-like stars---is the main goal of the PLATO mission. TESS mission (Ricker et al. 2016) planet yield simulations predict detections for about 10 HZ transiting planets suitable for atmospheric follow-up. All those targets will orbit bright (Ks$<$10),  mid-M dwarfs (Barclay et al. 2018). For directly imaged planets, taking the census of known M dwarfs within 20 pc from the Sun and adopting the occurrence rate from Dressing \& Charbonneau (2015), Crossfield (2016) estimates 5-10 rocky, HZ planets around nearby M dwarfs are resolvable by  instrumentation on ground-based Extremely Large Telescopes (ELTs), like the Giant Magellan Telescope  (GMT) and the Thirty Meter Telescope (TMT). The reflected-light planet-to-star contrasts for exo-Earths around M stars (10$^{-7}$--10$^{-8}$) are feasible with ELTs.  Contrasts for AFGK stars (10$^{-9}$--10$^{-11}$) are too steep for detecting Earth-sized HZ planets from ELTs in the foreseeable future.   However, such planets can be detected in thermal (10\,$\mu m$) emission around ${\sim}$10 targets (Quanz et al. 2015), including bright, well-known stars like Alpha Cen AB, $\tau$ Ceti, $\epsilon$ Eri, and Procyon, etc. Combining all these numbers, we expect to have a sample of 10-20 planets in the next decade suitable to searches for atmospheric biosignatures with ELTs.

\vspace{3mm}
\noindent {\bf  2.1~${\rm\bf O_2}$, ${\rm\bf H_2O}$, and ${\rm\bf CH_4}$ using Transiting Planets and Reflected Light from Directly Imaged Planets}
\vspace{1mm}

\begin{figure}
 \begin{subfigure}[]{0.5\textwidth}
  \includegraphics[angle=270, width=\textwidth, trim=0.9cm 1cm 0.45cm 3cm 0cm,clip=true]{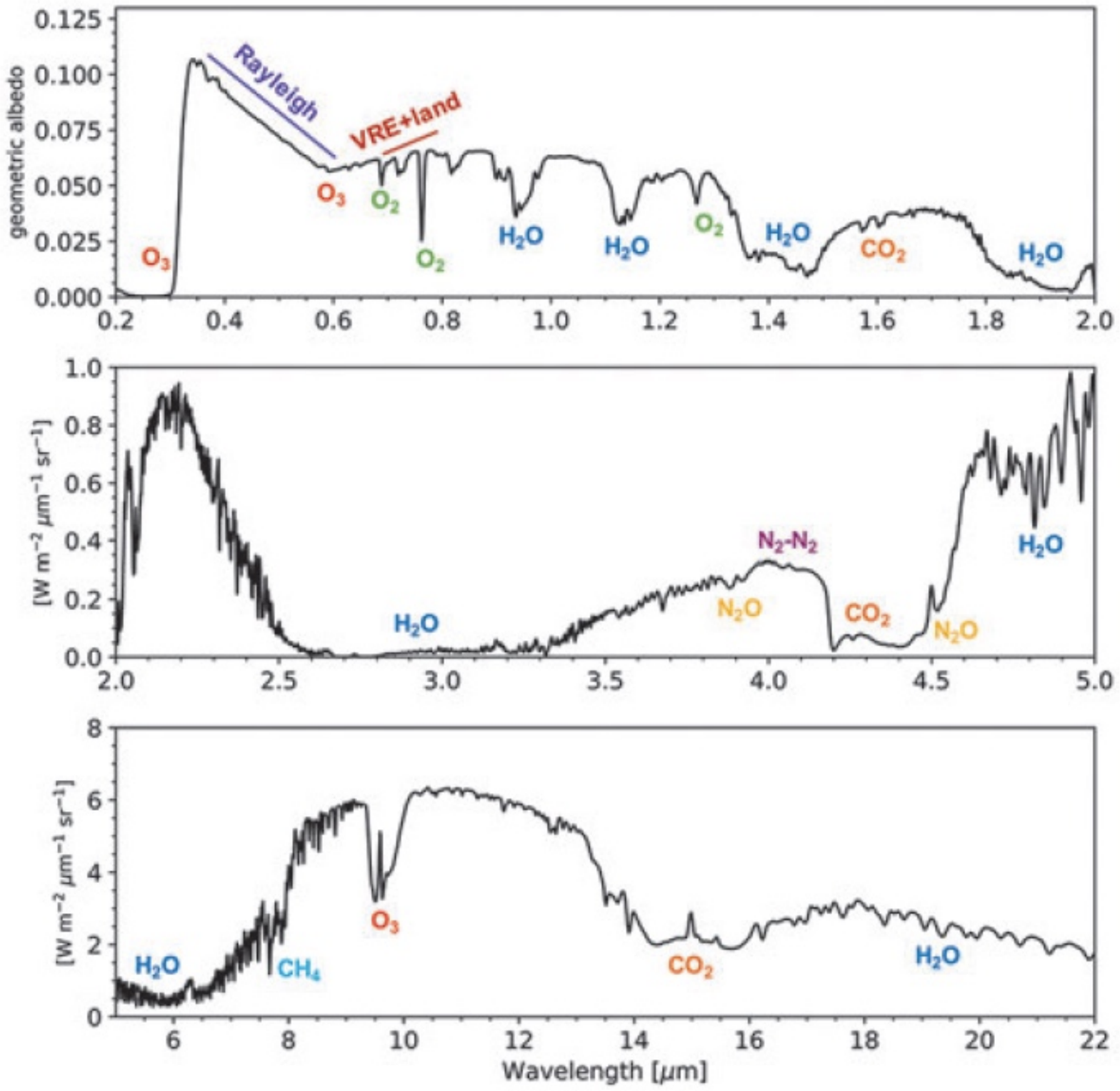} 
\label{fig:radiance}
\end{subfigure}
\begin{subfigure}[]{0.5\textwidth}
\centering
\includegraphics[angle=270, width=\textwidth, trim=4.05cm 0cm 5.5cm 2cm,clip=true]
{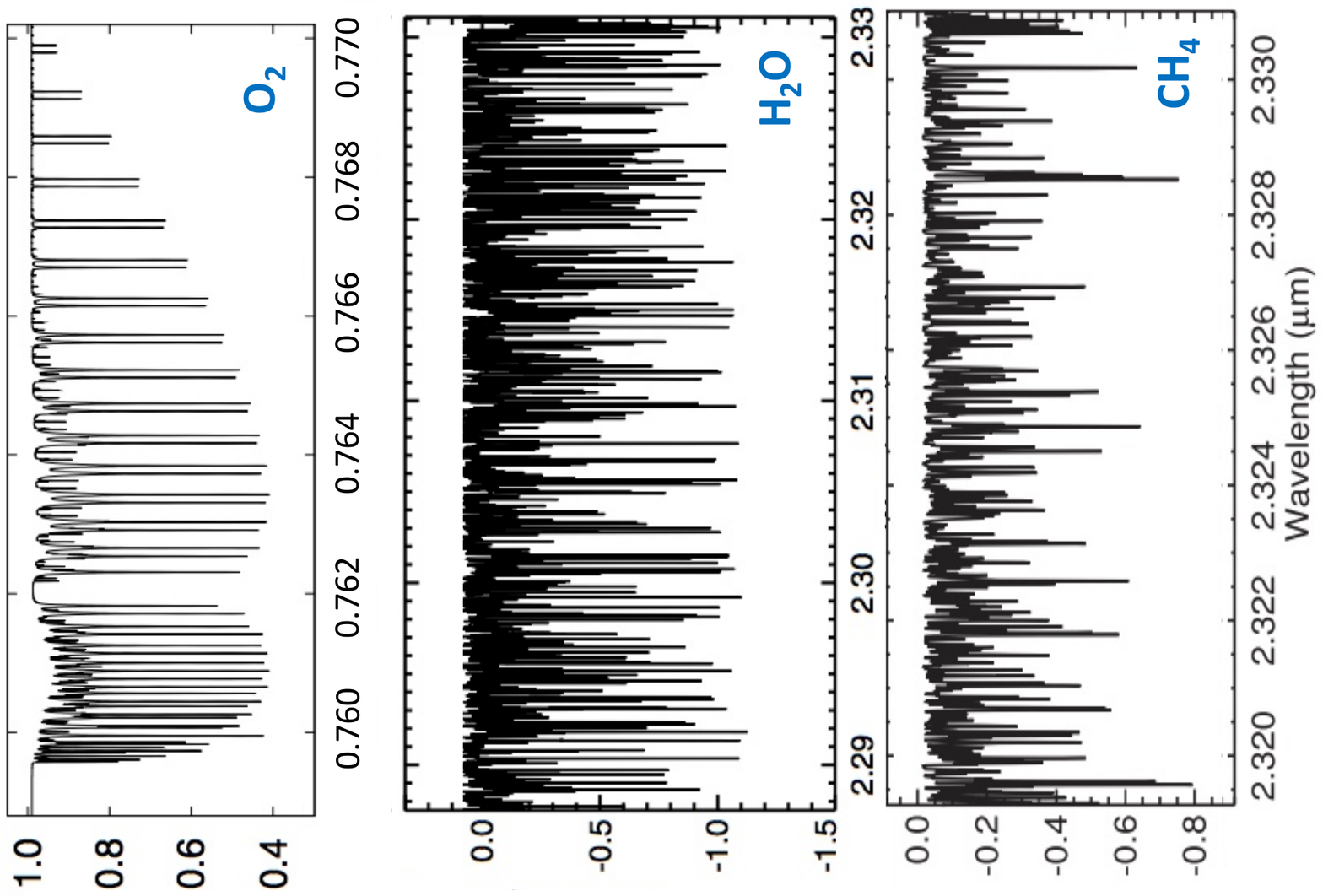}
\label{fig:molecules}
\end{subfigure}
\vspace{-0.21in}
\caption{{\it Left --} From top to bottom, model spectra of Earth biosignatures in the optical (transmission spectroscopy), near-IR (transmission spectroscopy and reflected-light imaging for M dwarfs), and thermal IR (direct imaging in thermal emission for AFGK stars). Figure from Schwieterman et al. (2018).  {\it Right --} Examples of absorption bands of ${\rm O_2}$, ${\rm H_2O}$, and ${\rm CH_4}$ at resolutions of R${\sim}$100,000, which is capable of resolving individual molecular lines.}
\label{fig:fig3}
\vspace{-0.165in}
\end{figure}

${\rm O_2}$, ${\rm H_2O}$, and ${\rm CH_4}$ produce prominent absorption bands at wavelengths shorter than ${\sim} 3.5\,\mu$m (Figure~\ref{fig:fig3}, left panel). ${\rm O_2}$ has well isolated, sharp bands at 0.688\,$\mu$m (B-band), 0.762\,$\mu$m (A-band), and 1.269\,$\mu$m ($a^1$ $\Delta$g band). 
${\rm H_2O}$ absorption features are ubiquitous throughout visible and infrared wavelengths, with prominent bands at 0.94\,$\mu$m, 1.13\,$\mu$m, 1.4\,$\mu$m, 1.9\,$\mu$m, and 2.3\,$\mu$m. ${\rm CH_4}$ has prominent bands near 2.3\,$\mu$m and 3.3\,$\mu$m, although they overlap with ${\rm H_2O}$\footnote{While JWST in principle could probe biosignatures via transit spectroscopy, those detections will be challenging.  
Due to large required telescope time allocations and expected noise floor limits, JWST will only be able to observe the atmospheres of a very small number of rocky planets.
Additionally, ${\rm O_2}$'s most prominent absorption appears at visible wavelengths for which JWST is not optimized.}.

These molecules can be detected by observing the transmitted spectrum of a transiting planet, or the reflectance spectrum of a directly imaged planet. For transits, the detectability of some molecules will be affected by their distribution as a function of atmospheric height, because of refraction (e.g. Betremieux \& Kaltenegger 2014), clouds or hazes. For example, on a planet with an Earth-like atmospheric chemical profile, transmission spectroscopy will miss  ${\rm H_2O}$:  ${\rm H_2O}$ in our atmosphere is mostly concentrated at the bottom 10--12\,km, which cannot be probed for planets in HZ of stars M4-5V or earlier. However, this might not be an issue for planets around smaller stars.  For reflected-light direct imaging, light from the host star that gets reflected from the planet will also have imprinted in it the signal of molecules present in the planetary atmosphere: near-infrared (0.9--2.5\,$\rm \mu m$) reflectance spectra are sensitive to detectable signatures of ${\rm O_2}$, ${\rm H_2O}$, and ${\rm CH_4}$. 

\vspace{3mm}
\noindent {\bf 2.2~ ${\rm\bf CH_4}$, ${\rm\bf O_3}$, and ${\rm\bf CO_2}$ using Thermal Emission from Directly Imaged Planets}
\vspace{1mm}

As shown in Fig.~\ref{fig:fig3} ({\it left}), a prominent ${\rm O_3}$ band appears in the middle of the $N$-band window (7-14 \,$\mu$m) accessible at dry sites like Maunakea and Las Campanas (9.6 \,${\rm \mu}$m).   ${\rm CO_2}$ absorption (15\,${\rm \mu}$m) overlaps with the filter's red edge.
${\rm CH_4}$ has an absorption feature at 7.7\,${\rm \mu}$m (see Fig.~\ref{fig:fig3}, {\it left}).
These biosignatures are broad, suitable for both spectroscopic and photometric observations. They may also help distinguish between rocky planet spectra indicative of a modern, habitable Earth or Earth in earlier evolutionary phases (e.g. an Archean Earth, Proterozoic Earth; Fig.~\ref{fig:archean}).  

\vspace{3mm}
\noindent {\bf 2.3~Detection of Biosignatures using High-Resolution Spectroscopy}
\vspace{1mm}

High-resolution spectroscopy, combined with cross-correlation techniques
(CCTs), can successfully detect molecules such as ${\rm CO}$, ${\rm H_2O}$, ${\rm TiO}$, ${\rm HCN}$, and metallic ions (Ti+, Fe+, Fe) in the atmospheres of gas giants (see e.g. Snellen et al. 2010, Brogi et el. 2012, Rodler et al. 2012, Nugroho et al. 2017, Hoeijmakers et al. 2018) from the ground. These same techniques can be perfected to detect  
biosignatures in transiting and directly imaged planets, as shown in ongoing simulation work (Snellen et al. 2013, Rodler \& L\'opez-Morales 2014, Snellen et al. 2015, Lovis et al. 2017, Wang et al. 2017,  Ben-Ami et al. 2018, Serindag \& Snellen 2019, L\'opez-Morales et al. submitted). 
Building instruments for ELTs optimized for biosignature detection is critical for the success of these observations (see e.g. Lovis et al. 2017, Ben-Ami et al. 2018). Complete molecular model line lists are also critical. In a high resolution spectrum, each molecule presents a characteristic pattern of absorption lines (Figure~\ref{fig:fig3}, {\it right}).  In addition to their  patterns, the lines from the exoplanet will appear Doppler-shifted with respect to their vacuum wavelengths by an amount equal to the relative velocity of the system with respect to Earth. By combining a template spectrum of the molecule we are searching for with information about the velocity of the system and CCTs, we can detect these molecules in the exoplanet. CCTs are more sensitive than low resolution observation methods traditionally used to search for molecular bands by a factor of ${\rm \sqrt{N}}$, where ${\rm N}$ is the number of resolved individual lines in the molecular band. However, for CCTs to be efficient, we need more complete molecular line lists (see e.g. Fortney et al. 2016, 2019).


\vspace{3mm}
\noindent {\bf 2.4~Feasibility}
\vspace{1mm}

The estimated amplitude of \otwo, \water, \cotwo, \ozone, and \methane\ bands in Earth-sized planets probed from the ground is of the order of ${\rm 10^{-5}}$ for transiting planets and of the order of ${\rm 10^{-7}-10^{-8}}$ for directly imaged planets. 
Current telescopes cannot detect those signals. 
As described in several test cases below, ground-based ELTs can detect biosignatures without unrealistically lengthy programs:

\textbf{Biosignature Detections from Transits} -- Simulations show that we will need 34 transits (or equivalently 70 hours) to achieve a $3\sigma$ detection of ${\rm O_2}$ at 0.76\,$\mu$m with planned instrumentation on the GMT for a planet with identical atmospheric composition to Earth's, orbiting in the HZ of a M4V star at 5 parsecs (Rodler \& L\'opez-Morales 2014). More recent calculations by Serindag \& Snellen (2019) give similar numbers. Combining observations with the GMT and TMT can cut that number of transits in half. The same observations with a 10-meter-class telescope would require over 200 transits (440 hours).

\textbf{Biosignature Detections from Reflected-Light Direct Imaging} --
From Kawahara et al. (2012), 1.5 hours of integration time achieving a planet-to-star contrast of 10$^{-8}$ could yield 5$\sigma$ detections of ${\rm O_2}$ at 1.27 $\mu$m  with IFS instrumentation on the TMT for an Earth analog orbiting an M star at 5 parsecs\footnote{Note that calibrating down the speckle halo to this limit may require more telescope time.  Achieving the required raw and post-processed contrasts, while challenging, is feasible when coupled with advanced methods for wavefront control, coronagraphy, and post-processing/spectral forward-modeling (Guyon et al. 2018; Currie et al. 2018). }. These observations are not possible with 8--10-meter-class telescopes 
because of the Inner Working Angles (IWA) imposed by diffraction limits (typically IWA ${\sim}$ 2--3\,$\lambda$/D) restrict planet detections to separations of about 0.1" or greater, regardless of whether 10$^{-8}$ contrast could be reached.   By comparison, the maximum projected separation for an Earth-twin (Earth-mass planet receiving an Earth-like insolation) around the nearest M stars (e.g. Proxima Cen, Barnard's Star)  is $<$ 0.06--0.1".  Simulations  suggest that, for reasonable expected performance, habitable Earth twins are detectable around at least 20 low-mass stars just from Maunakea (Currie 2018).

\textbf{Biosignature Detections from Thermal IR Direct Imaging} -- At 10 $\mu m$, exceptional image quality due to an extreme AO system (Strehl Ratio of $\sim$ 0.99) coupled with advanced coronagraphy (e.g. a charge-2 vortex) yields an extremely deep rejection of stellar halo light.   Detections are primarily thermal background limited.   As an example, under the driest feasible conditions on Maunakea or Las Campanas (detector-limited background) and assuming increased sensitivity from new detector technology (e.g. a Geosnap array), a 1.7 Earth-radius, 290 $K$ planet at the inner edge of the HZ around $\epsilon$ Eri (3.12 pc) could be detected at SNR = 32 in 30 hours (or $\sim$ 8 in two hours) over a broad 10--12.5 $\mu m$ bandpass and SNR $\sim$ 20 in 1 $\mu m$ narrowband filters out of the ozone absorption feature.   About 10 stars are suitable for detecting 1-2 Earth-radius planets at 10 $\mu$m in a few hours, which could justify deeper exposures to measure the depth of ozone.

\vspace{3mm}
\noindent {\bf 3. Relationship with and Complementarity to NASA Exo-Earth Detecting Missions}
\vspace{1mm}

The National Academy of Sciences {\it Exoplanet Science Strategy Consensus Study Report} recommended both 1) investment in ELTs and 2) a NASA mission to image Earth-like planets such as HabEx or LUVOIR, which has received strong community support (Plavchan et al. 2019).   Searches for Earth-like planets are also key scientific motivations for constructing ELTs that both complement and augment the scientific capabilities of these future NASA missions.  

The strategic importance of ELTs for finding Exo-Earths with NASA missions is described in a separate white paper (Currie et al. 2019).  Briefly, ELTs will likely achieve first-light well before HabEx/LUVOIR are launched.  Thermal IR imaging can identify candidate exo-Earths around AFGK stars, which could be followed up later with a NASA mission, providing biosignature detections (\ozone~ at 9.6 $\mu$m) and ancillary information (i.e. an estimate of the effective temperature, radius) crucial for better constraining atmospheric properties that would otherwise only be derived from optical reflectance spectra.   Under certain circumstances, rocky planets around the nearest M stars could also be followed up from space, also analyzed at complementary wavelengths.


\vspace{3mm}
\noindent {\bf 4.~Recommendations}
\vspace{1mm}

Along with understanding the universe's beginning and ultimate fate, the search for life on other planets and understanding the context for Earth are key focuses of modern astronomy.   Moreover, {\it are we alone} and {\it is Earth unique or rare} are not \textit{simply} scientific questions but more fundamentally are \textit{deeply human questions answerable by science}.  

ELTs can provide breakthroughs towards answering these questions, both complementing and laying the foundation for efforts with a future NASA Earth-finding mission.  To expedite efforts to characterize habitable rocky planets around nearby stars, we make the following recommendations the Astro2020 Decadal Survey Committee:

\begin{itemize}
\setlength\itemsep{0.1em}
  \item[1-] Include the search for Earth-like biosignatures on rocky planets around nearby stars as a key Decadal science case.
  \item[2-] Support the construction of the next generation of ground-based Extremely Large Telecopes (ELTs) over the next decade, which can provide the large apertures and spatial resolution required to start revealing the atmospheres of Earth-analogues around nearby stars. ELTs will complement and augument the science yield from future space-borne Earth-finding missions.
  \item[3-] Support the development of ground-based instrumentation optimized to detect biosignatures. 
  \item[4-] Support the generation of accurate line lists for potential biosignature gases, which are needed as model templates to detect those molecules.
\end{itemize}

\pagebreak
\noindent \textbf{References}

Barclay, T., Pepper, J., Quintana, E. V. 2018, ApJS, 239, 2

Ben-Ami, S. López-Morales, M., Garc\'ia-Mej\'ia, J., et al., 2018, ApJ, 861, 79

Betremieux, Y., Kaltenegger, L. 2014, ApJ, 791, 7

Brogi, M., Snellen, I. A. G., de Kok, R. J., et al., 2012, Nature, 486, 502

Brogi, M., Line, M. R. 2018, arXiv:181101681

Crossfield, I. J. M. 2016, arXiv:1604.06458

Currie, T., Brandt, T. D., Uyama, T., et al., 2018, AJ, 156, 291

Currie, T., 2018, {\it Breakthrough Science with the Thirty Meter Telescope}, Pasadena, CA

Currie, T., Belikov, R., Guyon, O., et al., 2019, Astro2020 White Paper {\it The Critical, Strategic Importance of Adaptive Optics-Assisted, Ground-Based Telescopes for the Success of Future NASA Exoplanet Direct Imaging Missions}

Domagal-Goldman, S. D.; Segura, A.; Claire, M. W., et al., 2014, ApJ, 792, 90

Dressing, C. D. \& Charbonneau, D. 2015, ApJ, 807, 45

Flowers, E., Brogi, M., Rauscher, E., Kempton, E.; Chiavassa, A. 2018, arXiv:181006099

Fortney, J. J., Robinson, T. D., Domagal-Goldman, S. et al. 2016, arXiv:1602.06305

Fortney, J. J., et al., 2019, Astro2020 White Paper, {\it The Need for Laboratory Work and Ab Initio Simulations to Aid in The Understanding of Exoplanetary Atmospheres}

Gaudi, S. et al. 2018, arXiv:180909674

Guyon, O., Mazin, B., Fitzgerald, M., et al., 2018, Proc. SPIE, 10703, 107030Z

Hoeijmakers, H. J., Ehrenreich, D., Heng, K., et al. 2018, Nature, 560, 453

Irwin, J. M., et al. 2015, 18th Cool Stars, Stellar Systems, and the Sun, 767–772

Jehin, E., Gillon, M., Queloz, D., et al.\ 2011, The Messenger, 145, 2 

Kaeufl, H.-U. 2004, SPIE, 5492, 1218

Kaltenegger, L., Traub, W. A., Jucks, K. W. 2007, ApJ, 658, 598

Kawahara, H., Matsuo, T., Takami, M., et al., 2012, ApJ, 758, 13

Kietavainen, R. \& Purkamo, L. 2015, Frontiers in Microbiology, 6, 725

Kopparapu, R.; Ramirez, R. M.; SchottelKotte, J., Kasting, J. F., et al. 2014, ApJL, 787, 29

Lafreniere, D., Marois, C., Doyon, R., Nadeau, D., Artigau, E., et al., 2007, ApJ, 660, 770

Lessner, D. 2009, Methanogenesis Biochemistry, doi:10.1002/9780470015902.a0000573.pub2

L\'opez-Morales, M., Haywood, R. D., Coughlin, J. L., et al., 2016, AJ, 152, 204 

Lovelock, J.E., 1965, Nature 207, 9-13

Luger, R. \& Barnes, R. 2015, AsBio, 15, 119

Marois, C., Correia, C., Galicher, R., et al., 2014, Proc. SPIE, 9148, 91480U

Martinache, F., Guyon, O., Jovanovic, N., te al. 2014, PASP, 126, 565

Meadows, V. S., 2006, Direct Imaging of Exoplanets: Sci \& Tech, Proc. IAU Col., 200, 25-34

Meadows, V. S., Reinhard, C. T., Arney, G. N., et al. 2018, AsBio, 18, 630

Nugroho, S. K., Kawahara, H., Masuda, K., et al. 2018, A\&AP, 612, A49 

Plavchan, P., et al., 2019, Astro 2020 White Paper, {\it Community Endorsement of the National Academies “Exoplanet Science Strategy” and “Astrobiology Strategy for the Search for Life in the Universe” Reports}

Quanz, S., Crossfield, I., Meyer. M., et al. 2015, IJAsB, 14, 279

Ricker, G. R., Vanderspek, R., Winn, J., Seager, S., et al. 2016, SPIE, 9904, 2

Rodler, F., L\'opez-Morales, M., Ribas, I. 2012, ApJL, 753, 25

Rodler, F., \& L\'opez-Morales, M.\ 2014, ApJ, 781, 54

Sagan, C., Thompson, W. R., Carlson, R., Gurnett, D., Hord, C., 1993, Nature, 365, 715

Schwieterman, E. W., Kiang, N. Y., Parenteau, M. N., et al. 2018, AsBio, 18, 663

Seager, S., Bains, W., Petkowski, J. J,. 2016, Astrobio, 16, 465

Serindag, D. B., Snellen, I. A. G., 2019, ApJL, 871, 7

Snellen, I. A. G., de Kok, R. J., de Mooij, E. J. W., Albrecht, S., 2010, Natur, 465, 1049

Snellen, I., de Kok, R., Birkby, J. L., 2015, A\&A, 576, 59

Swift, J.~J., Bottom, M., Johnson, J.~A., et al.\ 2015, JATIS, 1, 027002 

Wang, J., Mawet, D., Ruane, G., Hu, R., Benneke, B. 2017, AJ, 153, 183

Wolfgang, A., Rogers, L. A., Ford, E. B. 2016, ApJ, 825, 19

Wordsworth, R. \& Pierrehumbert, R. 2014, ApJ, 785, 20

\end{document}